%% file: main.tex
\documentclass{article}
\usepackage{ijcai21}
\usepackage{times}
\usepackage{helvet}
\usepackage{courier}
\usepackage{graphicx}
\usepackage{longtable}
\usepackage{xcolor}
\usepackage{tablefootnote}
\usepackage{times}
\usepackage{fancyhdr,graphicx,amsmath,amssymb}
\usepackage[ruled,vlined]{algorithm2e}
\usepackage[shellescape]{gmp}
\usepackage{multirow}
\include{pythonlisting}

\begin{document}
 
%
\title{Did Chatbots Miss Their 'Apollo Moment'\thanks{By {\em Apollo Moment}, we  refer to the opportunity for a technology to attain pinnacle of its impact.}? A Survey of the Potential, Gaps and Lessons from Using Collaboration Assistants During COVID-19}

\author{Biplav Srivastava \\
AI Institute, University of South Carolina
}

\maketitle
\begin{abstract}
 
Artificial Intelligence (AI) technologies have long been positioned as a tool to provide crucial data-driven decision support to people. In this survey paper, we look at how AI in general, and collaboration assistants (CAs or chatbots for short) in particular, have been used during a  true global exigency - the COVID-19 pandemic. The key observation is that chatbots missed their {\em Apollo moment} when they could have really provided contextual, personalized, reliable decision support at scale that the state-of-the-art makes possible.
We review the existing capabilities that are feasible and methods, identify the potential that  chatbots could have met, the use-cases they were deployed on, the challenges they faced and gaps that persisted, and  draw lessons  that, if implemented,  would make them more relevant in future health emergencies.

\end{abstract}

\input{introduction}
\input{background}

\input{chatbot-health}
\input{covid-usage}
\input{covid-gaps}
\input{future_guidance}

\input{conclusion}


\bibliographystyle{named}
\bibliography{main}

\end{document}

%% file: introduction.tex
\section{Introduction}

COVID-19\footnote{The virus named SARS-CoV-2, also called the novel corona virus, causes the COVID-19 disease. We will refer to the disease by COVID as well as COVID-19\cite{who-cornona} .} is a global pandemic which started in China in the winter of 2019 and has spread around the world with over 100 million cases and killing more than two million people by Feb 2021 \cite{who-cornona}. As the disease has progressed, new hot spots of the disease have emerged first in South-East Asia, then Europe and then in US, South America and South Asia. The disease has evolved and regions around the world have also switched their responses frequently while waiting for an effective vaccine to be developed and widely available for lasting cure. The impact of COVID pandemic has varied globally over geography and time, as measured by number of cases and deaths, depending on  demographics of the local population as well as the public health policies implemented in response. A compilation of resources can be found at \cite{covid-wiki}.

\begin{table*}
\begin{tabular}{cc}

\begin{minipage}{.5\textwidth}

\begin{enumerate}
\item Understanding the disease
   \begin{enumerate}
      \item Disease spread and simulation models
      \item Insights by visualization
    \end{enumerate}

\item Understanding impact on society
    \begin{enumerate}
        \item Understanding mental depression from social posts
        \item Assessing economic impact – job loss, industrial decline
        \item Effect on Supply Chain
        \item Assess risks
    \end{enumerate}
    
\item Observing disease in people
    \begin{enumerate}
        \item Fever detection via images
        \item Tracking people’s movement
    \end{enumerate}
    
    \end{enumerate}

\end{minipage}
  
& 

\begin{minipage}{.5\textwidth}

 \begin{enumerate}
\item Guidance for individual actions
    \begin{enumerate}
        \item Screening/ triage tools
        \item Guiding about government benefits
        \item Vaccine appointments and scheduling
    \end{enumerate}
\item Guidance for group-level actions
    \begin{enumerate}
        \item Models for when to open economy
        \item Contact tracing following an incident
        \item Matching producers and consumers to meet demand, reduce loss: food, medical supplies

    \end{enumerate}
\item Guidance for policy actions
    \begin{enumerate}
        \item Understanding impact of policy choices (e.g. lockdowns, travel restrictions)
        \item Design of economic interventions
        \item Fighting fake news
    \end{enumerate}
\end{enumerate}

  \end{minipage}

\end{tabular}
\caption{Emerging applications of  decision support  (AI) for COVID-19. Chatbots are appropriate for only a subset when interaction of the AI system  with people is needed (right column).}
\label{tab:covid-decsupport}
\end{table*}

In all aspect about this exigency, decision support is needed. AI community has responded to the need at molecular, clinical and societal scales \cite{covid-ai-mapping,rao-ai-covid,etzioni-ai-covid,ai-apps-covid,ml-covid-models,covid-wiki}.
Early in the pandemic, authors like \cite{rao-ai-covid,etzioni-ai-covid,ai-apps-covid} highlighted various scenarios where AI could help in tackling COVID19 as well as some of the potential pitfalls. The AI efforts were helped by different types of data being freely made available, calls for open collaboration \cite{collab-covid} and a sense of urgency. In Table~\ref{tab:covid-decsupport}, a sample of AI's potential application during COVID-19 is shown. They range from decisions to foster understanding of the disease and its impact to helping take actions for individuals, groups and the society at large.

Many of these AI potentials were indeed realized. In  \cite{covid-ai-mapping}, the authors cataloged significant application of machine learning between Jan1 - Aug1, 2020 for analyzing protein to aid disease detection and treatment (molecular scale), the analysis of patient data like images and conditions to improve patient care (clinical scale) and analysis of cases and social media to predict disease severity, understand mis-information and communicate effectively (societal scale). 

However, not many efforts lead to field-ready deployment of AI. In \cite{ml-covid-models}, the authors reviewed machine predictive and diagnostic machine learning models that were published and since revised twice. In their NeurIPS 2020 talk in December, they reported gaps including: ML models were often evaluated using the AUC metric (Area Under the Receiver Operating characteristic Curve) but this is not the measure helpful in practice, good performance on test data did not mean the model will do good in practice, there were replication issues, there was more need to share data, models and code, and the authors did not advise the models to be used in practice.

More generally, apart from creating decision support aids, it is also necessary to convey the insights to people and enable them to make better decisions. For example, consider the public health policy topic of of whether to require wearing of   masks or face covering. Its usage has been very controversial in the United States due to perceived impingement over individual freedom\cite{mask-opinion-nyt}. Many models have been built showing that mask wearing is effective. But how do we convey this information for maximal impact? In \cite{cmap-tool}, the authors used the method of Robust Synthetic Control to show that masks can be effective. But such methods were not deployed at scale to change people's behavior and save valuable lives.

It turned out that people were often more effective in helping each other via social media platforms. For example, on Reddit, people discussed and helped each other about unemployment benefits \cite{reddit-employment-nyt} and mental health \cite{reddit-health}. 
But decision-support tools did not develop at the scale expected for these opportunities when needed most.

In this survey paper, will focus on a specific form of AI, a collaborative assistant (CA), also known as a conversational interface (CI), chatbot or dialog system\footnote{Some researchers use the term chatbot exclusively for agents that perform chit-chat. Instead, we use the terms chatbot interchangeably for them to mean task-oriented collaborative assistants which is the focus of this paper.}.
We will look at how they are built, the capabilities they can provide and how, even before the pandemic, their benefits in health scenarios were unconvincing. 
Then, we will discuss the actual usage of chatbots during COVID followed by the gaps that were found.
We will see that the issues discovered pre-COVID may help contextualize the gaps and slow speed seen in the adoption of chatbot applications during COVID.
We then  conclude with what lessons can be learnt for using chatbots for a future pandemic. To our knowledge, this is the first systematic review of the effectiveness of chatbots during COVID-19 and what interventions are needed to make this technology more relevant for society's decision support needs.

%% file: background.tex
\section{Background}

In this section, we will give a background in chatbots and how they have been positioned to be valuable in health. This will help contextualize the challenges which were faced when using them for COVID.

\subsection{Collaborative Assistants}
\label{sec:chatbots}


A collaborative assistant (CA) \cite{dialog-intro} is an automated agent, whether physical like robots or abstract like an avatar, that can not only interact with a person, but also  take actions on their behalf and get things done. A simple taxonomy of interfaces we consider is shown in Table~\ref{tab:chatbot-types}. One can talk to a chatbot or, if speech is not supported, type an input and get the system's response. They can be embedded along with other interaction modalities to give a rich user experience. The chatbot may converse without a goal in pleasantries and hence not need access to data sources, or be connected to a static data source like a company directory or a dynamic data source like disease cases or weather forecast. The application scenarios become more  compelling when the chatbot works in a dynamic environment, e.g., with sensor data,  interacts with groups of people who come and go rather than only an individual at a time, and adapts its behavior to peculiarities of user(s). 
A few are shown in the right column of Table~\ref{tab:covid-decsupport}.

There is  a long history of CAs going back to 1960s when they first appeared to answer questions or do casual conversation\cite{dialog-intro}. 
In terms of conversation structure, a {\em dialog} is made up of a series of {\em turns}, where each turn is a series of {\em utterances} by one or more participants playing one or more {\em roles}. As examples, an on-line forum can have a single role of {\em users} while a customer support dialog may have the roles of {\em customer} and {\em support agent}. 
The most common type of chatbot deals with a single user at a time and conducts informal conversation, answers the user's questions, provides recommendations in a given domain and also takes actions on their behalf, if delegated. 
It needs to handle uncertainties related to human behavior and natural language, while conducting dialogs to achieve system goals. 



\begin{table}[t!]
\scriptsize
\centering
\begin{tabular}{|l|l|l|l|l|}
\hline
\textbf{S.No.} & \textbf{Dimension} & \textbf{Variety} \\ \hline
1 & User & 1, multiple \\ \hline
2 & Modality &  only conversation,  only speech, \\ 
  &   & multi-modal (with point, map, ...) \\ \hline
3 & Data source & none, static, dynamic \\ \hline
4 & Personalized &  no, yes \\ \hline
5 & Form & virtual agent, physical device, robot \\ \hline
6 & Purpose & socialize, goal: information seeker,  \\ 
  &  & goal: action delegate \\ \hline
7 &  Domains & general, health, water, traffic, ...\\ \hline

\end{tabular}
\caption{Different Types of Conversation Interfaces}
\label{tab:chatbot-types}
\end{table}



\subsection{Building Data-Consuming Chatbots}
\label{sec:builders}

The core problem in building chatbots is that of dialog management (DM), i.e., creating dialog responses to user's utterances.
Given the user's utterance, it is analyzed to detect their intent and a 
policy for response is selected. This policy may call for querying a database, and the result is returned which is used by response generator to create a response using templates. The system can dynamically create one or more queries which involves selecting tables and attributes, filtering values and testing for conditions,  and assuming defaults for missing values. It may also decide not to answer a request if it is unsure of a query's result correctness.

Note that the DM may use one or more domain-specific data bases (sources) as well as one or more domain-independent sources like language models and word embeddings. 
When the domain is dynamic, 
the agent has to execute actions to monitor the environment, model different users engaged in conversation over time and track their intents, learn patterns and represent them, reason about best course of action given goals and system state, and execute conversation or other multi-modal actions. As the complexity of DM increases along with its dependency on domain dependent and independent data sources, the challenge of testing it increases as well.


There are many approaches to tackle DM  in literature including finite-space, frame-based, inference-based and statistical learning-based \cite{chatbot-survey-statistical-ml,chatbot-book,minim-dialog,young2013pomdp}, of which,  finite-space and frame-based are most popular with mainstream developers. Indeed, commercial chatbots have popularized a frame-based approach where the domain of conversation like travel booking is organized into dialog states called frames (like flight booking) which consists of variables called slots, their values and prompts to ask user (for the values). An example of slot is origin of a flight that the user wants to book.

Task-oriented DMs  have traditionally been built using rules for selecting frames and slots, with some learning to identify user's intent. 
Further, a DM contains several independent modules which are optimized separately, relying on huge amount of human 
engineering .
The recent trend in research is to train DM from end-to-end, allowing error signal from the end output of DM to be back-propagated to raw input, so that the whole DM can be jointly optimized \cite{e2e-dialog-learning}.



\subsection{Discussion: Implementation Choices, Evaluation and Fairness Issues with Chatbots}

Given the plethora of implementation methods,
recent surveys for building chatbot are \cite{chat-building-survey} where the  authors summarize the different approaches for building conversation systems  and identify challenges, and  \cite{dl-chatbot-survey} which focuses on deep-learning based methods for building chatbots. 
There is renewed interest in inference-based methods to control DM behavior \cite{cohen-2019-foundations,dialog-planning-adi,dialog-planning-muise}.
In  \cite{chatbot-arch-icse18}, the authors look at requirements and design options to make chatbots customizable by end users as their own personal bot \cite{chatbot-arch-icse18}.

There are ongoing efforts to evaluate chatbots as well. Prominent are Dialog System
Technology Challenge (DSTC), a series of competitions whose ninth edition happened in 2021\cite{dstc9-overview}. Each competition has multiple tracks to benchmark chatbots automatically based on various interaction and problem solving capabilities.  Another competition is ConvAI \cite{convai-chatbot-competition} which evaluates conversations based on human evaluation of dialog quality. 

The emerging consensus in the dialog  community is that while the current approaches, especially deep learning based approaches, are effective in building increasingly engaging
chatbots for simple scenarios with clear goals and in the presence of large training data, more research is needed to build systems which are collaborative problem solvers \cite{allen-chatbot-complexconv,tyson-aaai18,tyson-ic,cohen-2019-foundations} and can control behavior \cite{dialog-planning-adi}.
Such systems deal with iteratively refined goals,  need the ability to reason about evolving information and domain, and adds unique value when the chatbot can take a pro-active role in dialog when it is confident of completing a task with available information.

Further, like much of AI, chatbots are data-driven and have been known to have issues like implicit bias when using pre-trained domain-independent models, prone to adversarial attack, potential sources of privacy violations, safety concerns and abusive language \cite{dialog-ethics}. 
Addressing them is an area of active research \cite{chatbot-rating,chatbot-recipes}.


%% file: chatbot-health.tex
\section{Chatbots in Health and their Performance (Pre-COVID)}
\label{sec:chatbot-health}

Chatbots have been built for health applications from the very beginning of dialog research as even the first system, Eliza \cite{eliza},  simulated a  Rogerian psychotherapist. In a 2018 survey \cite{chatbot-health-review}, the authors conducted a meta-review of papers on evaluation of conversational agents in health on major digital libraries until 2018. They found 
 that  of the 14 chatbots matching their inclusion criteria of robust use, more than half of the systems were built for self-care. 
 The most common strategy for dialogue management was finite-state (6) and frame-based (7); deep-learning based systems were not prominent. 
 
 They also found that empirical evaluation for chatbots was not as rigorous as other technologies in health since the gold-standard methods like randomized controlled trial (RCT) were not common and patient safety was rarely evaluated in those studies. In only one study, RCT established the efficacy of a conversational agent (Woebot) to have a significant effect in reducing depression symptoms (effect size d = 0.44, p = .04).

 
In another 2018 study of \cite{chatbot-harm}, the authors had conducted a small experiment where 54 subjects were asked to use commercial chatbot systems (from Amazon, Apple and Google) for medical help and their experiences were analyzed. The participants were only able to complete 168 (43\%) of the assigned 394 tasks. Of these, 49 (29\%) reported actions that could have caused  harm, including nearly half - 27 (16\%) - of deaths. Looking carefully at the chat transcripts, one could notice that the systems were making errors in understanding the users' request (intent) or they were giving narrow factual answers which the users could misinterpret  as medical recommendation in the context of their overall task.

In another study from 2020 \cite{chatbot-harm-2}, the authors considered the performance of 8 commercial systems (from Amazon, Apple, Google, Microsoft and Samsung) on questions  (prompts) related to what authors called safety-critical scenarios (e.g., violence, mental health) and lifestyle (e.g., diet, smoking). 3 people evaluated 240 responses to 30 prompts. Responses were manually evaluated along a rubric that checked characteristics of the systems response like the user's intent was identified. A response to  safety-critical question was deemed appropriate if it included a referral to a health professional or service, while a response to lifestyle question were deemed appropriate it provided relevant information to address the problem raised. The authors found that the systems collectively responded appropriately to 41\%(46/112) of the safety-critical and 39\% (37/96) of the lifestyle prompts.


\subsection{Discussion}

The long history of using chatbots in health would have suggested that the technology would be effective in achieving better health outcome. However, existing studies did not establish this even before COVID.
Although the studies differed 
in their specific design and findings about available commercial chatbots, they indicated a general inappropriateness to handle medical queries without oversight. 

In this context, a  white paper appeared from World Economic Forum \cite{chatbot-reset} in late 2020 that provides a framework for how chatbots should be developed for health applications.
It identifies that the key stakeholders, apart from users, are health service providers (chatbot operators), developers and regulators. The framework identifies steps that the stakeholders can take so that a chatbot can be useful, exhibit competency and build trust with users.


%% file: covid-usage.tex
\section{Potential for Collaborative Assistants during COVID}

As the COVID pandemic started, there was a rush to build chatbots for various scenarios. 
For example, a May 2020 study reported that public health organizations deployed systems around four main scenarios  \cite{chatbot-usage}:  
    (a) share information and triage patients
    (b) monitor symptoms
    (c) support for behavior change 
     (d) support for mental health.  
Later, more usages appeared like universities guiding students on campuses\cite{univ-chatbot-covid} and agencies scheduling vaccine appointments. 
In  Table~\ref{tab:covid-decsupport}, among the AI application areas, chatbots were used for those involving direct action by individuals whereas they could have been helpful for more. 

We now look at these usages in detail under the  category of share information, monitor symptoms and  provide support. 

\subsection{Share Information}

The first wave of chatbots shared information about COVID19 \cite{chatbot-covid-wef}. For example, the 
World Health Organization (WHO) provided resources to alert people around the world using messaging platforms \footnote{ https://www.who.int/emergencies/diseases/novel-coronavirus-2019/, https://www.who.int/news-room/feature-stories/detail/who-health-alert-brings-covid-19-facts-to-billions-via-whatsapp}.
However, while valuable, they were offering simple, generic question answers. 

In the US, a report \cite{us-chatbot-covid-usage} of June 2020 noted that three-fourth of US states were developing chatbots to disseminate information about COVID19 and unemployment benefits to residents since there was an upsurge in (customer service) calls for information to government agencies. General guidelines and best practices emerged for building such chatbots with focus on public health \cite{health-chatbot-covid} and children \cite{pediatric-health-chatbot-covid}.
Institutions of higher education also started planning deployment to answer common student questions
\cite{univ-chatbot-covid,univ-chatbot-covid-nyt}. Since unemployment grew, chatbot like BEBO \cite{app-bebo} were built to share information about unemployment benefits. 

\subsection{Monitor symptoms, Triage Patients and Guide for Treatment}

COVID19 pandemic also triggered many regions to launch mobile and web based digital assistants to guide people when they should take medical assistance. 
One of the most common usage was triage, i.e., determining which potential patients should seek urgent medical care. In \cite{chatbot-triage}, the author describes how hospital facilities are using chatbots built using commercial platforms to screen patients. Chatbots were also used to allow residents to self-report conditions with the aim to collect data and help public health authorities  in the United Kingdom\footnote{https://covid.joinzoe.com/us-2}.

At the national level, Singapore launched the TraceTogether app \footnote{https://www.tracetogether.gov.sg/} for monitoring people and alerting them when others with suspected cases may have have come in their contact or vice-versa.
India launched the Aarogya Setu mobile app to self-report health condition and track vulnerable persons to give alerts when they may have come in contact with suspected cases. A study into its working and experience \cite{india-contact-tracing} reported that the tool using Bluetooth and Global Positioning System (GPS) is effective but there are security concerns.  
India is planning to use an app, called CoWin, to guide people on when they can get the COVID vaccine \footnote{https://gadgets.ndtv.com/apps/news/cowin-app-registration-download-application-modules-2347477}. 

At a smaller scale of  campuses, many
universities and companies  planned to use mobile apps to track well-being of their occupants \cite{univ-chatbot-covid-nyt,univ-chatbot-covid}.
One of the first in the US was CovidWatch \cite{app-covidwatch}. But their adoption was slowed down by concerns about user privacy and liabilities \cite{univ-covid-app-privacy}.

\subsection{Supporting Residents and Customers}

COVID19 accelerated the deployment of chatbots for customer service applications in businesses
\cite{jobs-chatbot-covid}. While the benefit of chatbots in reducing a companies costs is clear since they will be substituting  existing manpower by technology, its benefit to the customer is unclear. In fact, the competency of chatbots has been in question leading some businesses to advertise access to human agents as a competitive differentiator.

COVID also accelerated usage of chatbots which provide support to people with mental health issues  \cite{mental-health-chatbot-covid}. One of them, Woebot, had been found to positively useful even before COVID \cite{chatbot-health-review}.
However, despite their popularity, it is experimentally unclear if any of the tools provided substantive or better support than human providers during COVID.

\subsection{Discussion}

COVID triggered launching of new chatbots that were specific to the disease, its impact (e.g., on employment and education) as well as accelerated adoption of existing chatbots in customer care and mental health. 
However, most of them had a narrow focus, could answer simple questions but were not collaborative or complex problem solvers, were not personalized, could not handle group usage, and left open questions about  usability, effectiveness and
handling of user privacy.

%% file: covid-gaps.tex
\section{Gaps Found in Using Chatbots During COVID}

In this section, we identify some of the major gaps discovered during chatbot deployment for COVID.

\subsection{Inconsistent Ability (G1)}


Users found COVID chatbots to handle simple questions well but struggle with complex ones. 
 An test early in the pandemic found that for the same condition, different chatbots created by different institutions, but claiming to be compliant to the guidelines of US's Center for Disease Control (CDC), would give opposing results for the same condition \cite{chatbot-covid-varying-response}.
Another  study surveyed participants for whether they would trust chatbots provided by reputable organizations \cite{covid-chatbot-user-reaction}. Here, trust refers to the ability of the chatbot to answer the question, integrity to perform what it committed to (if any), and benevolence by keeping patient interest in focus. 
The authors
found that users are neutral to who provides them COVID information - humans or chatbots - as long as the latter is competent in answering the queries. 

\subsection{Missing Differentiation  Over Alternatives (G2)}

Users often had multiple alternatives (website, phone lines) to get information and there was no compelling need just to use a chatbot. Further, the capability of chatbots was limited and users needs were left unmet\cite{covid-chatbot-user-reaction}.

\subsection{Inaccessible Information (G3)}

A majority of the chatbots created assumed that the users knew English, were literate (could read and write), were savvy with digital devices (like smartphones) and did not have disabilities. These assumptions left out (or delayed rollout to)  a significant section of the society around the world which could have been avoided because work on digital inclusiveness predates COVID.

\subsection{Ambiguity Regarding User Privacy (G4)}

Contact tracing apps and chatbots proposed for COVID need access to a mobile phone user's location and connectivity resources like Bluetooth. Prominent phone vendors like
Google and Apple built interfaces to allow Bluetooth contact tracking using Android and iPhone devices but regions around the world were concerned about how user's data was stored and processed. 
In one study \cite{chatbot-privacy-law}, the authors noted that digital surveillance contributed to the success of certain countries (China, Singapore, Israel, and South Korea) in controlling cases. The authors observe that during uncertain times of the pandemic, having expansive regulatory clarity like General Data Protection Regulation (GDPR) was an advantage for system design that is compatible with human fundamental rights but in contrast, having a patchwork of narrow rules like {\em "US Health Insurance Portability and Accountability Act (HIPAA), and even the new California Consumer Privacy Act (CCPA), leave gaps that may prove difficult to bridge in the middle of an emergency"}.

Even at the smaller scale of campuses, many
universities and companies  who planned to use mobile apps to track well-being of their members and visitors found resistance due to concerns over perceived invasion of individual's privacy \cite{univ-covid-app-privacy}.

\subsection{Insufficient User Testing (G5)}

The field of testing for chatbots is still in inception \cite{chatbot-testing-tips,chatbot-testing}. Further, in the rush to release systems quickly, testing of COVID chatbots was not sufficient as evidenced in reported behavioral disparities \cite{chatbot-covid-varying-response}. This affects the perceived trustworthiness of the information given by a chatbot and reflects negatively on the organizations developing it.

\subsection{Discussion}

Users found COVID chatbots to have limited capability (e.g., handle simple questions well but struggle with complex ones), have inconsistent behavior and not sufficiently tested. Users also had concerns about privacy of their data and system being safe or trustable. 


%% file: future_guidance.tex
\section{Lessons for a Future Exigency}

Based on the experience of chatbots during COVID and gaps discovered, we now identify some lessons that, if implemented, would make chatbots more helpful in a future health exigency.

\subsection{Identify Key Value to Provide With Chatbots}

A key question to ask, when someone is developing a chatbot, is why it is needed over any other alternative available. The best use-cases are those where no alternative is suited more than the chatbot's unique property that it is a sequential modality for interaction in natural language with the conversation evolving based on a user's inputs.
Such a focus will also address gap (G2).

In many scenarios, the interaction between the agent and user does not need to  be sequential (e.g., the user knows what they want at the outset), the user does not care about interacting in natural language (e.g., can enter a structured input like phone number or zipcode) and the system 
can use multiple modalities to show results.
For example, to find the nearest hospital, an alternative to a chatbot can be a webpage where the user can give the full request  if they already know it (current location), and get the result (address and directions)  in just one interaction. 

A chatbot should be used in a scenario when it will add value, preferably uniquely, to the user.
Given the health setting, a list of such  scenarios can be compiled. Some examples are: when the topic is sensitive (e.g., mental health), the subject is new (e.g., vaccine), legal record of interaction has to be maintained for possible audit. 
One can also create frequent questions and articulate how their answers help meet business benefits desired from the chatbot.

\subsection{Create Health Chatbot Development Best Practices}

There is a need to develop best practices for the health domain and meet the gaps G1, G3, G4 and G5.

\noindent {\bf Methodology for Chatbot Testing}:
Testing of software for meeting requirements and usability is a challenging endeavor. For chatbots, they pose additional challenges since the behavior of the system is dependent not just on DM algorithms but also on data procured for development  and user's inputs and history of conversation.
Some approaches \cite{chatbot-testing} and checklist \cite{chatbot-testing-tips} have emerged and more are needed. Furthermore, existing ones will have to  be  customized for health applications in line with regulations for data privacy \cite{chatbot-privacy-law} and electronic devices in that domain
 \cite{fda-devices}.

\noindent {\bf Guidance on Data Handling and Privacy:} As noted in \cite{chatbot-privacy-law,univ-covid-app-privacy}, ambiguity regarding data privacy emerged as a barrier towards adoption of chatbots during COVID. 
An emerging framework for health chatbots, Chatbot RESET
 \cite{chatbot-reset}, launched in late 2020, provides guidance on how developers, health service providers and regulators can navigate the space.
 It consists of a set of AI and ethics principles as applicable for health use-cases of chatbots and then makes recommendations along the dimensions of optional, suggested and required based on risk to a patient.

\noindent {\bf Guidance on Regulations and Medical Liabilities:}
In health regulations, the role of medical devices and the liabilities it creates for different stakeholders is well understood \cite{fda-devices}. However, the same is not clear for chatbots. Depending on the criticality of health scenario involved, chatbots need to characterized so that they can be appropriately developed, tested and transparently marketed to users \cite{chatbot-reset}. This will spur development of trustable, secure and reliable chatbots.

\subsection{Chatbot Generators}

Once the design and content of a chatbot is unambiguous, it should be possible to automatically generate it for many usability factors like language, conversation style, color schemes and multi-media modalities. This ideas was proposed in 
 \cite{chabot-generator} for chatbots consuming Open Data but the idea is general purpose. This will help meet G3.

\subsection{Making Chatbots Trustable}

There are many promising efforts that can help meet G4 and G5.
In \cite{chatbot-recipes}, the authors discuss how to handle trust issues with chatbots to make them safe. The broad approaches are (a) Unsafe Utterance Detection which involves training and deploying additional classifiers for detecting unsafe messages, (b) Safe Utterance Generation which
involves training the model such that it is unlikely to produce unsafe content at run time, (c) Sensitive Topic Avoidance which involves avoiding sensitive topics, and (d) Gender mitigation strategies where the model  is forced to  respond  with  gender  neutral  language. The method needs to be adapted for health.

In \cite{chatbot-rating}, the authors propose an approach to test and rate chatbots from a third-party perspective for trust using customizable issues like abusive language and information leakage. Such ratings can help in making chatbots more acceptable to users especially in mental health applications.


%% file: conclusion.tex
\section{Conclusion}

COVID19 caused  a major disruption in the lives of people around the world and  they were looking for help with decisions in all aspects of their lives. At this juncture, chatbots as the AI technology for providing personalized decision support at scale, was most needed. In this paper, we reviewed the range of methods available to build them and capabilities they can offer. We then looked at how chatbots were positioned for benefit in health and the limited evidence that existed before COVID of their impact. COVID triggered launching of disease-specific new chatbots as well as accelerated adoption of existing one in customer care and mental health. 
However, most of them worked in simple scenarios
and raised  questions about  usability, effectiveness and handling of user privacy.
We identified gaps from the experience and  drew lessons that can be used for future health exigencies. 
